\documentclass{aastex}
\input psfig.sty
\def\today{\ifcase\month\or jan\or 
  feb\or mar\or apr\or may\or jun\or
  jul\or aug\or sep\or oct\or nov\or dec\fi
  \space\number\day, \number\year}
\shortauthors{G. Walker, E. Shkolnik, D. Bohlender, S. Yang }
\shorttitle{ RV precision of fiber-fed spectrographs }

\begin{document}

\title{The Radial Velocity Precision of Fiber-fed Spectrographs}

\author {Gordon A. H. Walker\altaffilmark{1}, Evgenya Shkolnik\altaffilmark{1}}
\affil{Physics \& Astronomy Dept., UBC, Vancouver BC, Canada V6T 1Z1} 

\altaffiltext{1}{Visiting Astronomer, Canada-France-Hawaii Telescope,
operated by the National Research Council of Canada, the Centre
National de la Recherche Scientifique of France, and the University of
Hawaii.}

\author {David A. Bohlender\altaffilmark{1}} \affil{National Research
Council of Canada, Herzberg Institute of Astrophysics\\ 
Victoria BC, Canada V9E 2E7}

\author {Stephenson Yang}
\affil{Physics \& Astronomy Dept., UVic, Victoria BC, Canada V8W 3P6}

\begin{abstract} 

We have measured the radial velocities of five 51 Peg-type stars and
one star known to be constant in velocity. Our measurements, on 20
\AA\/ centered at 3947 \AA, were conventional using Th/Ar comparison
spectra taken every 20 or 40 minutes between the stellar exposures.
Existing IRAF routines were used for the reduction. We find
$\sigma_{RV}$ $\leq$ 20 m s$^{-1}$, provided 4 measurements (out of 72)
with residuals $>5\sigma_{RV}$ are neglected. The observations were
made with the CFHT Gecko spectrograph (R$\sim$110,000), fiber-fed by
the CAFE system.  $\sigma_{RV}$ $\leq$10 m s$^{-1}$ seems possible with
additional care. This study was incidental to the main observing
program and is certainly not exhaustive but the small value of
$\sigma_{RV}$ implies that the fiber feed/image slicer system on Gecko
+ CAFE, essentially eliminates the long standing problem of guiding
errors in radial velocity measurements. We are not promoting this
conventional approach for serious Doppler planet searches (especially
with Gecko which has such a small multiplex gain), but the precision is
valuable for observations made in spectral regions remote from telluric
lines or captive-gas fiducials. Instrument builders might consider the
advantages of the CAFE optics which incorporate agitation and invert
the object and pupil to illuminate the slit and grating, respectively,
in future spectrograph designs.

\end{abstract}


\section{Introduction} 

Radial velocity measurements are of fundamental importance in
astronomy. For stars, absolute radial velocities can be uncertain by as
much as a km s$^{-1}$ because of motions in the stellar photosphere.
Differential velocities, on the other hand, can be much more precise
yielding important information about pulsation for single stars and
masses for those in double or multiple systems. The most spectacular
return from high precision radial velocities in the last decade has
been the detection of the minute reflex accelerations caused by unseen
planetary companions for some 100 solar-type stars (see, for example,
\cite{Mar00}).

The basic design of spectrographs most commonly used to determine
stellar Doppler shifts has changed little in over a century.  The
telescope images a star on the spectrograph slit which is collimated,
dispersed and then reimaged onto a detector.  The principal
improvements have been in detectors, from eye to photographic plate to
photomultiplier or solid state array. In modern spectrographs gratings
largely replace prisms for dispersion.  Precision in estimating
wavelength displacements is compromised by motion and defocus of the
star image at the slit. Both translate into a displacement of the
stellar spectrum relative to the comparison spectrum. Flexure of the
spectrograph decollimates Cassegrain instruments while, for
bench-mounted spectrographs, flexure of the telescope causes its
optical axis to wander in the spectrograph. Both effects lead to
spectral shifts.

\cite{PF67} thoroughly studied these effects with the McKellar
spectrograph at the coud\'{e} focus of the DAO 1.2-m telescope from
photographic spectrograms. Working at a resolution of $\sim$55,000,
they found that even for bright stars with well modulated spectra,
there were external errors of nearly 300 m s$^{-1}$, while for the sky
they were closer to 100 m s$^{-1}$, and internal errors were even lower
at $\sim$70 m s$^{-1}$. They pointed out that without some scrambling
of starlight at the slit, such systematic errors would remain. The
subsequent introduction of pupil or image slicers reduced this
discrepancy by improving the uniformity of slit illumination.
\cite{Cam79}, also used the McKellar spectrograph to demonstrate that,
with a captive-gas to impose wavelength fiducials directly in the
stellar spectrum, systematic errors could be dramatically reduced.
Nowadays, with iodine vapor cells, errors have been reduced to a few m
s$^{-1}$, the level at which precision begins to be limited by the
natural velocity noise of the star's atmosphere \citep{Mar00}.

For many programs it is not always possible or appropriate to include
either suitable telluric lines (generally water vapor) or captive-gas
lines such as I$_{2}$. This is certainly true for the blue/ultra-violet
at high spectral resolution. In this paper we briefly demonstrate that,
with minimal precautions, a precision of $\leq$20 m s$^{-1}$ is
possible for R$>$100,000 with a fiber-fed spectrograph over runs
lasting at least several nights and probably for much longer.

We became aware of this high precision following an observing run at
CFHT in 2001 to observe Ca II H \& K in the spectra of 51~Peg-type
stars. These stars had been discovered to have Jupiter-mass secondaries
with orbital periods of a few days from precise radial velocity
measurements, mostly using captive gas fiducials. Such high quality
radial velocity standards were not available to \cite{PF67} but offered
us an excellent test of the Gecko + CAFE ({\bf CA}ssegrain {\bf F}iber
{\bf E}nvironment) radial velocity precision. In a run one year later
we deliberately took comparison arc spectra before and after every
stellar spectrum and it is these spectra which we discuss here.

Our data were certainly adequate to calculate improved periods for each of the stars. The revised periods and phases will be published in another paper.

\section{The Observations}

\subsection{The spectra}

The observations were made on 26 to 30 July 2002 UT with the Gecko
\'{e}chellette spectrograph fiber fed by CAFE \citep{Bau00} from the
Cassegrain focus of the Canada France Hawaii 3.6-m telescope (CFHT).
Spectra were centered at 3947 \AA\/ in the 14th order which was
isolated by a UV grism with some 60 \AA\/ intercepted by the CCD. The
dispersion was 0.0136\AA\/ pixel$^{-1}$ and the 2.64 pixel FWHM of the
thorium-argon (Th/Ar) lines corresponded to R=110,000. The spectra were part of a long
term program to monitor variations in the Ca II H \& K reversals of
stars with short periods (3 to 4.5 days) planets. For this reason, they
were of high signal-to-noise in the continuum. The detector was a
back-illuminated EEV CCD (13.5 $\mu$m$^{2}$ pixels) with spectral
dispersion along the rows of the device.  Single Th/Ar
arcs were taken immediately before and after each stellar spectrum and,
like the flat fields, were fed through the same fiber as the starlight.
Probably more comparison spectra should have been taken but, as time
was of the essence for this program, only single comparison spectra
were taken.

Fiber modal noise was suppressed by continuously agitating the fiber
close to its output (see \cite{Bau01}). A Fabry lens at the fiber
output projects the pupil as input to the spectrograph while
simultaneously illuminating the grating with the object field.
\cite{Bau98} have pointed out that inversion of object and pupil
improves both spectral resolution and spectral stability. In this
arrangement the image projected on the grating is the fiber core output
aperture enlarged 1000 times, while the fiber far field image is
transformed into a pseudo slit by a four slice Bowen-Wallraven 
slicer of silicon which projects to some 50 pixels length in the spectrum.
For this program there was no on-chip binning of pixels.
 
Because the CCD has a number of pixels that suffer from
non-linear dark signal, we took a large number of dark exposures with
integration times matching those of the various stellar, Th/Ar, and
flat-field exposures, and created an average dark for each exposure
time.  Then, rather than using conventional biases to remove the
baseline from each observation in the data reduction, the appropriate
mean darks were subtracted from the stellar, Th/Ar, and flat-field
exposures.  Flat-fields were then normalized to a mean value of unity
along each row and the exposures for all stars observed on a night were
combined into a mean object to define a single aperture for the
extraction of all stellar and comparison exposures, including
subtraction of residual background between spectral orders (prior to
flat-fielding).  This aperture was ultimately used to extract one
dimensional spectra of the individual stellar and comparison  exposures
and a one dimensional extraction of the mean, normalized flat-field
(without an interorder background subtraction).  The extracted stellar
and comparison exposures were then divided by the one dimensional
flat-field to obtain flat-fielded spectra.

\begin{figure}
\centerline{\psfig{figure=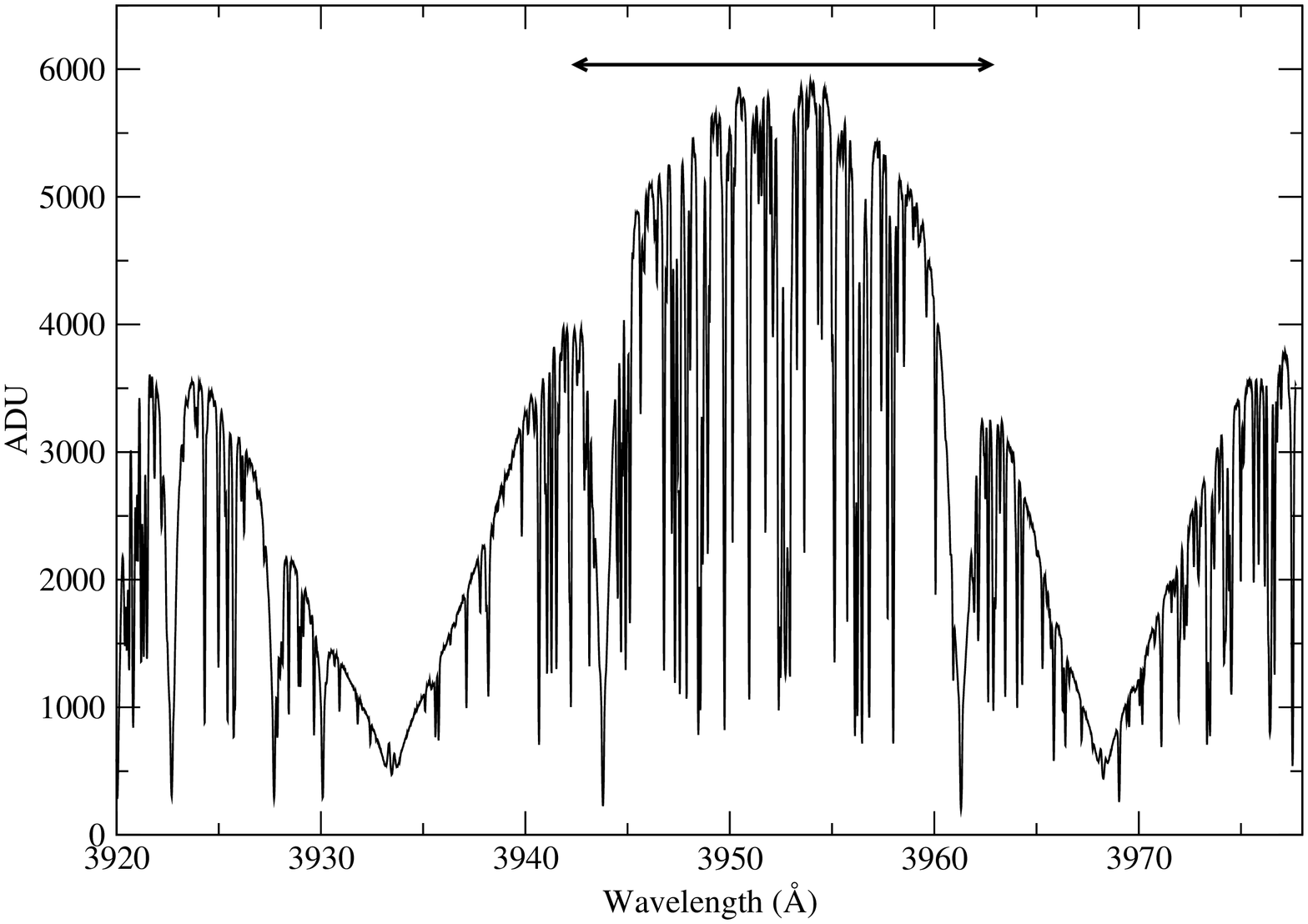,width=18cm,angle=-90}}
\caption{A single spectrum of $\tau$ Ceti flat-fielded as described in the text.  The arrows define the 3942-3963 \AA\/ region bounded by two strong lines of Al which was used in measuring the differential radial velocities ($\Delta$RV) discussed in the paper. 
\label{figure1} }
\end{figure}

A specimen, flat-fielded spectrum of $\tau$ Ceti is shown in Figure
\ref{figure1} with the $\sim$ 20 \AA\/ of the spectrum used to
measure the radial velocities indicated.

\subsection{The radial velocities}

Table \ref{table} lists the five program stars and the standard, $\tau$ Ceti,
plus spectral types and U magnitudes, the planetary orbital periods and velocity amplitudes, with details of the observations such as average S/N and exposure
time. 

Radial velocities were estimated with the {\bf fxcor} routine in
IRAF\footnote[1]{IRAF is distributed by the National Optical Astronomy
Observatories, which is operated by the Association of Universities for
Research in Astronomy, Inc. (AURA) under cooperative agreement with the
National Science Foundation.}. The Th/Ar comparison lines were used to
provide dispersion-corrected stellar spectra. A Fourier
cross-correlation was carried out on the dispersion-corrected spectra
taking the first spectrum in the series for each star as the template.
Hence, all differential radial velocities ($\Delta$RV) are relative to
the first spectrum on the first night.  Both the template and the
spectrum being measured were normalized with a low order polynomial and
the correlation was taken over that part of the spectrum bounded by (and
including) two strong aluminum lines ($\sim$3942-3963 \AA).

\begin{deluxetable}{cccccrccl}
\tabletypesize{\footnotesize}
\tablecaption{ The Program Stars \label{table} }
\tablewidth{0pt}
\tablehead{   
\colhead{~ star} & 
\colhead{spectrum} & 
\colhead{U} &
\colhead{exp time} &
\colhead{S/N\tablenotemark{a}} &
\colhead{\bf n\tablenotemark{b}} &
\colhead{K$_{max}$\tablenotemark{c}} &
\colhead{$\sigma_{RV}$} &
\colhead{P$_{orb}$\tablenotemark{d} ~} 
\\
\colhead{} &
\colhead{} &
\colhead{} &
\colhead{S} &
\colhead{} &
\colhead{} &
\colhead{m s$^{-1}$} &
\colhead{m s$^{-1}$} &
\colhead{days ~}  
}
\startdata
$\upsilon$ And&F7 V&4.69&1200&490&8&~70&11&4.6170\tablenotemark{e}\\
$\tau$ Cet&G8 V&4.43&1200&620&10&$-$&9&$-$\\
$\tau$ Boo & F7 VI &5.02&1200&470&12&467 & 31 (18)\tablenotemark{f} &3.3128\\
HD 179949&F8 V&6.83&2400&250&13&102&18&3.093\\ 
51 Peg&G2 IV&6.36&1200&270&17&~56&26 (18)\tablenotemark{f}&4.2293\\
HD 209458&G0 V&8.38&2400&150&12&~87&32 (18)\tablenotemark{g}&3.524738\\
 \enddata
\tablenotetext{a}{average per 0.0136\AA\/ pixel in the continuum}
\tablenotetext{b}{number of spectra}
\tablenotetext{c}{published values}
\tablenotetext{d}{published orbital period: $\tau$ Boo and $\upsilon$ And \citep{But97}, HD 179949 \citep{Tin00}, 51 Peg \citep{Mar96}, HD 209458 \citep{Cha99}}
\tablenotetext{e}{shortest of three planetary periods}
\tablenotetext{f}{value in brackets omits a single outlying point - see text}
\tablenotetext{g}{value in brackets omits two outlying points - see text}
\end{deluxetable}

\begin{figure}
\centerline{\psfig{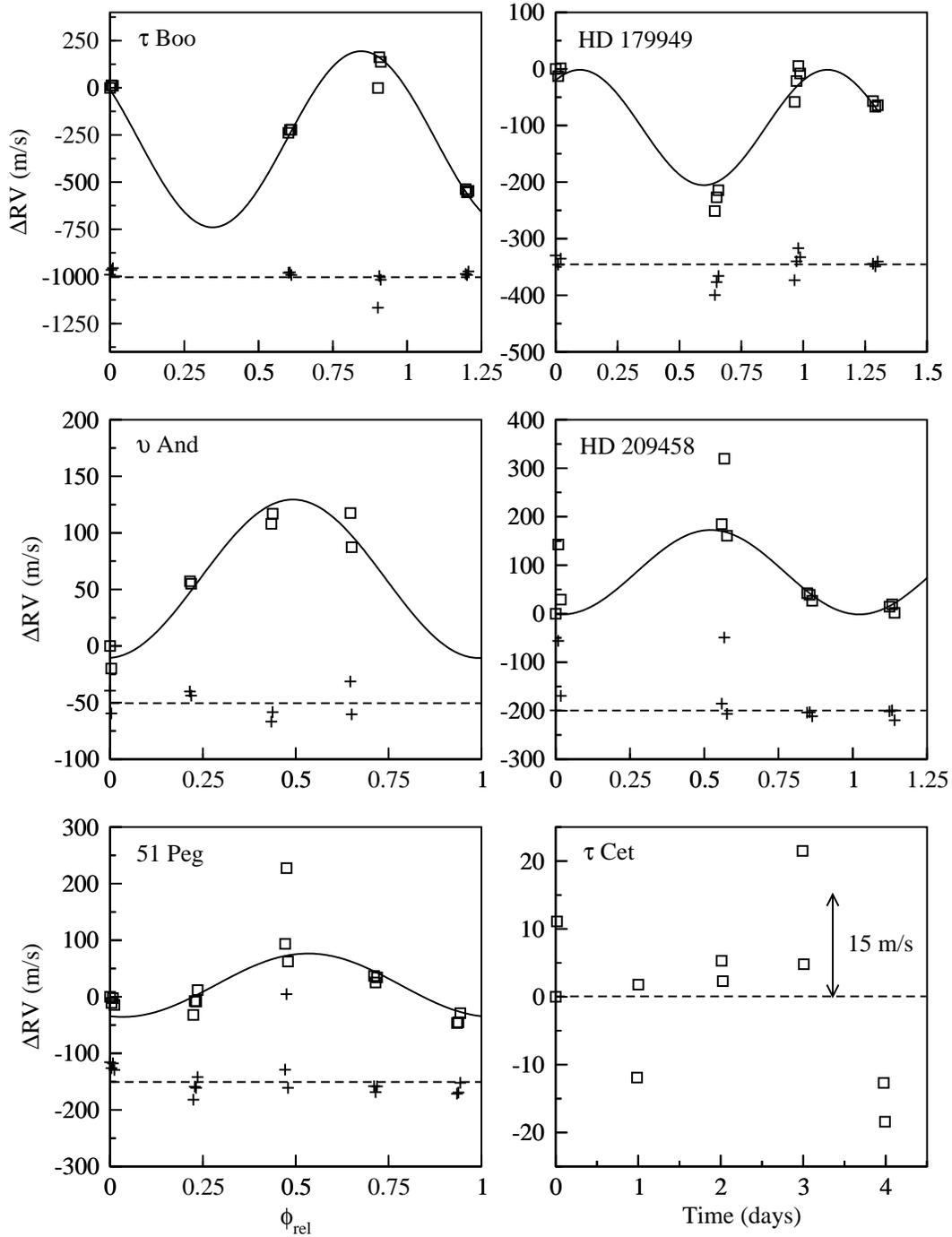}}
\caption{The differential radial velocities for the five 51 Peg stars
are plotted as a function of phase, and the unvarying star, $\tau$
Ceti,  as a function of time. The sine curves have the published
planetary orbital period and K$_{max}$ for each star (see Table
\ref{table}) and have been shifted in phase and $\Delta$RV to give the best fit to the
$\Delta$RVs. For $\tau$ Ceti there is a best fitting line of
constant velocity. Below each velocity curve are shown the residuals of
the $\Delta$RV from the curves.
\label{figure2} }
\end{figure}

In Figure \ref{figure2} the $\Delta$RVs for the five 51 Peg stars are
plotted as a function of relative phase, and for $\tau$ Ceti as a function of
time. The sine curves have the published planetary orbital period and
K$_{max}$ from Table \ref{table} and were shifted in phase and $\Delta$RV to give the
best fit. For $\tau$ Ceti there is a best fit is to a line of constant
velocity. Below each velocity curve are shown the $\Delta$RV residuals
from the curves. Note that, because the amplitude of the stellar reflex
velocity is very different in each case, the individual $\Delta$RV
scales vary widely.  Values of $\sigma_{RV}$ corresponding to the
residuals are listed in Table \ref{table}. If four out of the 72 data
points with $\sigma_{RV}>$100 m s$^{-1}$ are omitted, all of the
$\sigma_{RV}$ are $<$20 m s$^{-1}$ and the two stars observed at the
highest S/N have $\sigma_{RV}$ $\sim$10 m s$^{-1}$.

\subsection{The four extra-large residuals}

\begin{figure}
\centerline{\psfig{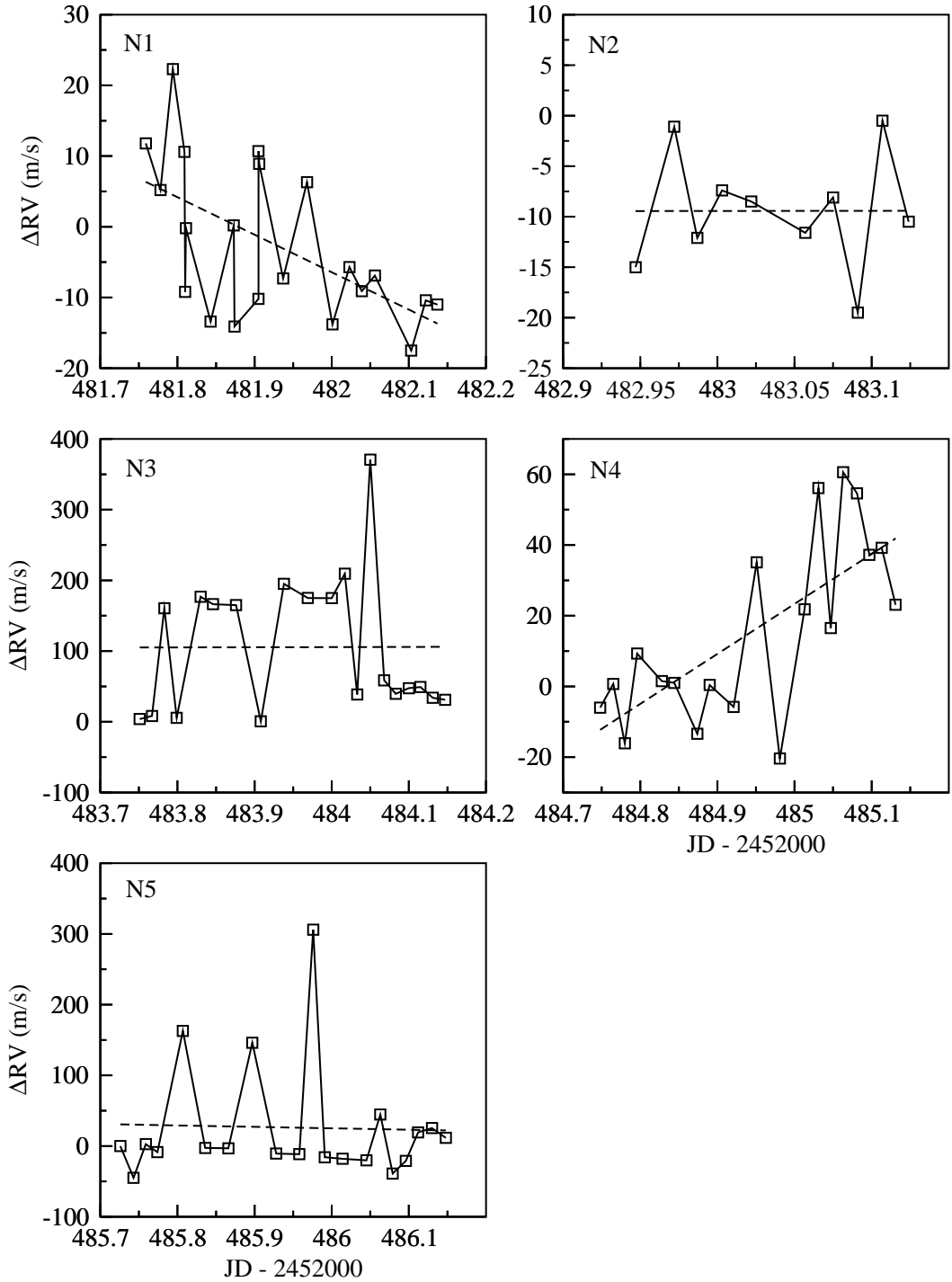}}
\caption{The shifts expressed as $\Delta$RV determined for the Thorium Argon comparison arc lines as described in the text for each night. Note that the velocity scales on nights 3 and 5 are an order of magnitude greater than on the other three nights.  The dashed lines delineate trends.
\label{figure3} }
\end{figure}

The motions of the comparison spectra expressed as $\Delta$RV on each
of the five nights are shown in Figure \ref{figure3}. The results are
surprising. The $\Delta$RV scales on nights 3 and 5 are an order of
magnitude greater than on the other three nights.  Nights 1 and 4 show
a trend and modest scatter, with RMS about the trend being 9 and 18 m
s$^{-1}$, respectively. Night 2 (only partially clear) shows no trend
and an RMS scatter of 9 m s$^{-1}$. Nights 3 and 5 have RMS scatters of
96 and 83 m s$^{-1}$, respectively!

The most likely reason for spectral shifts on the detector are liquid
nitrogen boil-off from the CCD Dewar, distortion of the coud\'{e} room
floor (Earth tides, dome stress etc.), and spectrograph `seeing'
effects. If these changes are slow and linear as they appear to be on
nights 1, 2 and 4, then interpolation of the comparison line positions
should properly calibrate the shifts for the stellar spectra.  Most of
the shifts are remarkably small considering that $\Delta$RV=10 m
s$^{-1}$ corresponds to 1.3$\times10^{-4}$ \AA\/ at 3947 \AA\/ which is
10$^{-2}$ of a pixel or 0.13 $\mu$m.  Monitoring the exposure meter to
estimate the centroid of the stellar exposures would have allowed more
accurate barycentric velocity corrections but we did not have such a
system to hand.

One might expect that the large, erratic residuals on nights 3 and 5
would be reflected in large reflex residuals for the stellar radial
velocities on those nights but this is not always so. The delinquent
velocity for 51 Peg on night 3 (see Figure \ref{figure2}) corresponds
to the very large comparison residual ($>$300 m s$^{-1}$).  By
contrast, the velocity for $\upsilon$ And observed immediately
afterwards which uses the same comparison spectrum agrees within a few
m s$^{-1}$ with the second value for $\upsilon$ And on that same
night. 

On the face of it, there is no consistent reason to reject the four
large residuals. They may represent `creaking' in the system which
introduces occasional uncalibratable jumps. Unfortunately, because of
the nature of the program we were not able to take comparison spectra
more often than every 20 or 40 minutes. In two cases, the comparison
lines show a residual of more than 100 m s$^{-1}$ which is reflected as
an almost equally large residual in the stellar velocity. In these
cases (51 Peg and HD 209458), using a mean Th/Ar which ignores the bad
points restores the stellar velocities to the curve. In the other two
cases ($\tau$ Boo and HD 209458) there is no problem with the Th/Ar
comparison.

Some of the large residuals may be inherent to the reductions and not
the  spectrograph. We used the same version of IRAF on two different
platforms, PC-IRAF V2.11.3 and SUN/IRAF V2.11.3. The former generated
several very large residuals in the Th/Ar comparison positions which
disappeared on reduction on a SUN.  The data presented in this paper
were all reduced on a SUN but it is still possible that some of the
large residuals are artifacts of the reduction. In short, we cannot be
sure that occasional large residuals are not a risk with such
conventionally determined velocities without a more extensive study.

Differential variations between the strenghts of comparison arc lines can
also limit precision. The temperature of the arc, its age, `on' time,
voltage stability etc., cause relative line strength variations and
introduce line shifts especially when lines are unresolved blends. To
judge from the low scatter in the comparison spectral shifts and values
of $\sigma_{RV}$ $\sim$ 10 m s$^{-1}$ for the two stars observed at
high S/N, variations in the arc lines has not been important above this
level and this $\sigma_{RV}$ may be the best achievable for single
spectra with this spectrograph, but a more careful study is necessary.

\section{Conclusions}

The effectiveness of fiber agitation in overcoming modal noise has
already been demonstrated \citep{Bau01}. In this paper we believe we
have demonstrated that the optics of the CAFE fiber system which invert
the conventional object and pupil \citep{Bau98} illumination of slit
and grating together with agitation effectively eliminates the large
guiding and tracking errors which once plagued conventional radial
velocity measurements. Our experience suggests that comparison arcs
taken with sufficient frequency are adequate to track slow spectrograph
drifts for a bench mounted fiber-fed instrument in a stable
environment. The precision depends on the resolution which, at 110,000
for Gecko, is better than 10 m s$^{-1}$. Being an \'{e}chellette
format, the spectral curvature is small, but an \'{e}chelle format
would provide a multiplex gain of one or two orders of magnitude over
the single spectra captured by Gecko (in this paper we only used 20
\AA!). From the point of view of radial velocities, an \'{e}chelle
would be much more sensitive.

Instrument builders might want to consider the advantages of CAFE in
future spectrograph designs. From our observations alone we cannot say
much about the stability of such conventional radial velocities over
periods of years but it will be interesting to follow it up. 

Conventional radial velocities even of the precision reported here are
not a substitute for those made with captive-gas fiducials in serious
Doppler planet searches but the precision is important for observations
made in spectral regions remote from telluric lines or captive-gas
fiducials such as we have described here.

\acknowledgements

Research funding from the Canadian Natural Sciences and Engineering
Research Council (G.A.H.W. \& E.S.) and the National Research Council
of Canada (D.A.B.) is gratefully acknowledged. We are also indebted to the
CFHT staff for their care in setting up the CAFE fiber system and the
Gecko spectrograph.

\clearpage

\clearpage

\begin{thebibliography}

\bibitem[Baudrand et al.(1998)]{Bau98} Baudrand, J., Jocou, L., Guinouard, I.,  1998, ASP Conference Series, Fiber Optics in Astronomy III, Vol. 152, 32.

\bibitem[Baudrand \& Vitry(2000)]{Bau00} Baudrand J. \& Vitry, R.,
2000, \procspie,  4008, 182.

\bibitem[Baudrand \& Walker(2000)]{Bau01} Baudrand J. \& Walker, G.A.H., 2001, \pasp, 113, 851.

\bibitem[Butler et al.(1997)]{But97}Butler, P., Marcy G., Williams, E., Hauser, H., Shirts, P., 1997 \apjl, 474, L115.

\bibitem[Campbell \& Walker(1979)]{Cam79}Campbell, B., \& Walker, G.A.H., 1979, \pasp, 91, 540.

\bibitem[Charbonneau et al.(1999)]{Cha99}Charbonneau, D., Brown, T., Latham, D., Mayor, M., Mazeh, T., 1999, \apj., 529, 45.

\bibitem[Marcy et al.(1996)]{Mar96}Marcy, G., Butler, P., Williams, E., Bildsten, L., Graham, J., 1996, \apj, 481, 926.

\bibitem[Marcy \& Butler(2000)]{Mar00} Marcy, G.W., \& Butler, R.P., 2000, \pasp, 112, 137.

\bibitem[Petrie \& Fletcher(1967)]{PF67} Petrie, R.M., \& Fletcher, J.M, 1967, I.A.U. Symposium {\bf 30}, 1967, (Ed: A.H. Batten \& J.F Heard, Academic Press), 43.

\bibitem[Tinney et al.(2000)]{Tin00}Tinney, C., Butler, P., Marcy, G., Jones, H., Penny, A., Vogt S., Apps, K., Henry, C., 2000, \apjl, 551, L507.

\end{thebibliography}
\end{document}